\documentclass[reprint,amsmath,amssymb, aps]{revtex4-2}
\usepackage{graphicx}
\usepackage{dcolumn}
\usepackage{bm}
\usepackage{hyperref}
\usepackage{xcolor}
\usepackage{caption}
\usepackage{subcaption}

\def\br{{\bf r}}
\def\rmd{{\rm d}}
\def\kB{{k_B}}
\def\g#1{{g^{(#1)}}}

\def\mathu{{\bm{\mathcal{U}}}}

\begin{document}
\title{Entropy approximations for simple fluids}
\author{Yang Huang}
\affiliation{
  Physics Department, Carnegie Mellon University, Pittsburgh PA, 15213
}
\author{Michael Widom}
\affiliation{
  Physics Department, Carnegie Mellon University, Pittsburgh PA, 15213
}
\date{\today}

\begin{abstract}
 Liquid state entropy formulas based on configurational probability distributions are examined for Lennard-Jones fluids across a range temperatures and densities. These formulas are based on expansions of the entropy in series of $n$-body distribution functions. We focus on two special cases. One, which we term the ``perfect gas'' series, starts with the entropy of an ideal gas; the other, which we term the ``dense liquid series'' removes a many-body contribution from the ideal gas entropy and reallocates it among the subsequent $n$-body terms. We show that the perfect gas series is most accurate at low density, while the dense liquid series is most accurate at high density. We propose empirical interpolation methods that are capable of connecting the two series and giving consistent predictions in most situations.
\end{abstract}
\maketitle

\section{Introduction}

{\em Ab-initio} phase diagram prediction depends on the ability to calculate free energies of competing phases. The Helmholtz free energy $F(N,V,T)=E-TS$ balances the energy $E$ against the entropy $S$. Electronic density functional theory (DFT~\cite{HohenbergKohn,KohnSham,VASP2} calculates the energy for any given atomic configuration. Entropy proves more difficult to calculate, yet it, too, can in principle be calculated for a given atomic configuration. The basis for this claim lies in the equivalence of the thermodynamic entropy with the Shannon Information~\cite{Shannon1948} required to specify the configuration. This recognition led us to calculate the substitutional entropies of high entropy alloys~\cite{Widom16,KimWidom}, vibrational entropies of solids~\cite{e24050618}, and the entropies of liquid metals~\cite{gao2018information,Widom2019,huang2022ab}.

Calculating thermodynamic properties of dense fluids such as liquid metals is especially challenging because they lack regular crystalline order, yet they are more strongly interacting than a vapor. HS Green~\cite{HSGreen1952}, following Kirkwood's factorization of $n$-body canonical ensemble correlation functions~\cite{Kirkwood1942}, introduced a series expansion for entropy starting from the $1$-body term,
\begin{equation}
  \label{eq:S1_DL}
  s_1^{DL}=\frac{3}{2}-\ln(\rho\Lambda^3),
\end{equation}
where $\Lambda=\sqrt{h^2/2\pi m k_BT}$ is the quantum thermal De Broglie wavelength, $\rho=N/V$ is the density, and we report entropy in units of $k_{\rm B}$. The $3/2$ term in~(\ref{eq:S1_DL}) arises from momentum fluctuations; the logarithmic term is the information required to specify the position of one atom in volume $V=1/\rho$ subject to spatial resolution $\Lambda$~\cite{e20070514}. Subsequent terms in Green's series, in the form of spatial integrals of logarithms of correlation functions, represent corrections to the entropy due to the information content of the correlations~\cite{Stratonovich1955,Widom2019}. We will name this series the ``dense liquid'' (DL) series, because we find it is most accurate at high densities.

Nettleton and MS Green~\cite{Nettleton1958}, followed by others~\cite{Yvon1969,Raveche1971,Mountain1971,Wallace1987,Baranyai1989,Hernando1990,Laird1992}, re-expressed the entropy in a series of grand canonical correlation functions starting with the exact entropy of an ideal (perfect) gas,
\begin{equation}
  \label{eq:S1_PG}
  s_1^{PG}=S^{IG}\equiv \frac{5}{2}-\ln(\rho\Lambda^3).
\end{equation}
We will call this series the ``perfect gas'' (PG) series because we find it is most accurate at low densities and $s_1^{PG}$ becomes exact as $\rho\to 0$. Note that
\begin{equation}
  \label{eq:S1}
  s_1^{PG}=s_1^{DL}+1.
\end{equation}
The additional $1$ arises from the term ``$-N$'' in the Stirling expansion of $\ln{N!}$~\cite{e20070514,Widom2019}) and thus may be considered as a many-body contribution related to particle interchange. The subsequent terms in the perfect gas series fall into two categories: one set that we shall refer to as ``compressibility terms'' integrate polynomials of $n$-body distribution functions; the other set that we refer to as ``information terms'' integrate logarithms of distribution functions.

\begin{figure}
  \includegraphics[width=.45\textwidth]{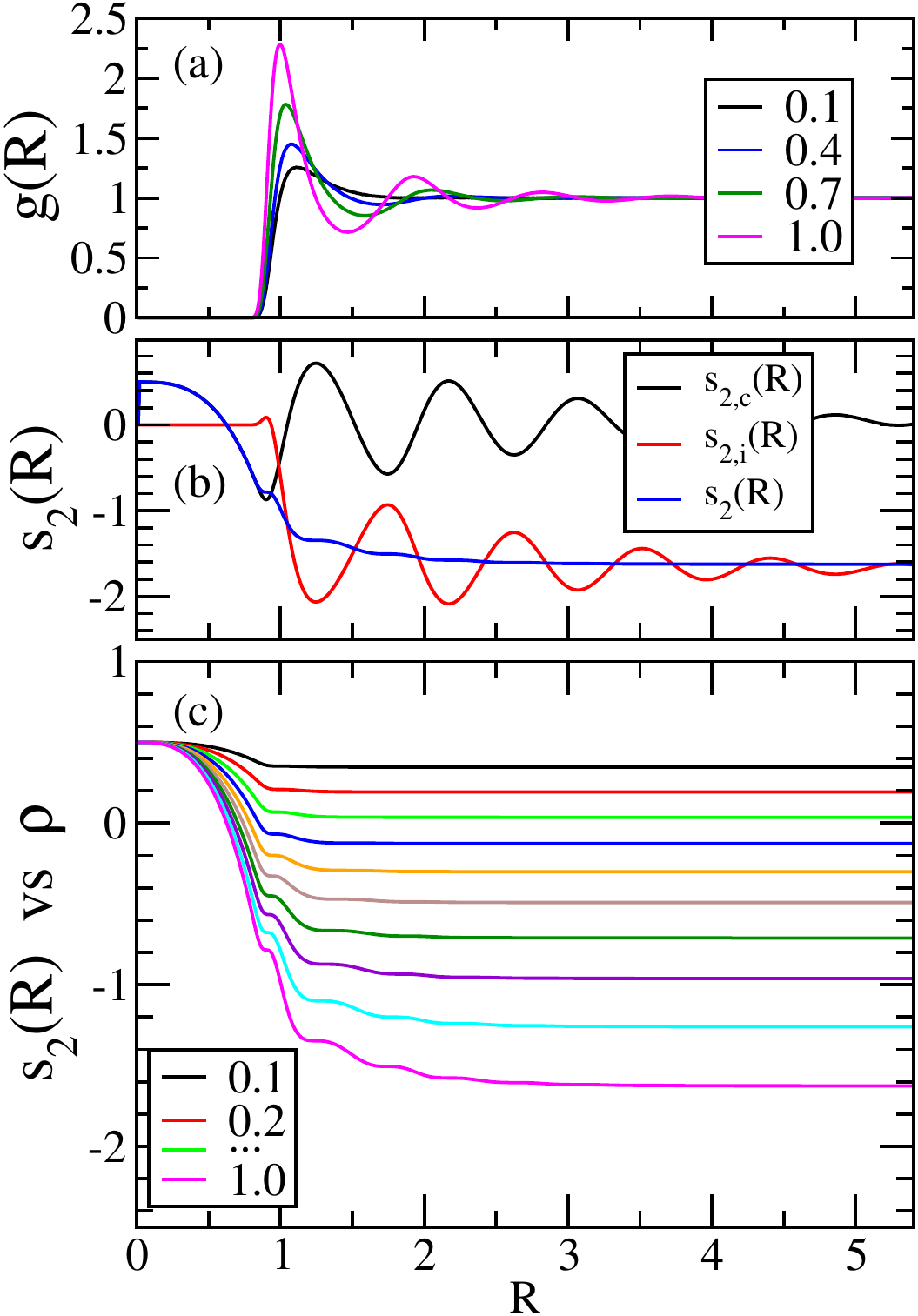}
  \caption{\label{fig:g2} (a) Pair correlations $\g{2}(r)$ at several densities and temperature $T=5$. (b) Convergence of the integrals for $s_2^{DL}$ as functions of the upper integration cutoff radius $R$. (c) Two-body entropies for all densities at T=5.}
\end{figure}

Baranyai and Evans~\cite{Baranyai1989} clarified the relationship between the DL and PG series. The extra $1$ contained in $s_1^{PG}$ can be reallocated among the $2$-body, $3$-body and higher terms according to the series representation
\begin{equation}
  \label{eq:expand_1}
  1 = \frac{1}{1\cdot 2} + \frac{1}{2\cdot 3} + \dots,
\end{equation}
and these terms exactly compensate additional compressibility terms that are polynomial in canonical ensemble correlation functions. They further point out that the combination of information and compressibility terms cancel long-range oscillations of the correlation functions (see Fig.~\ref{fig:g2}), at each order, leading to well-localized and ensemble-independent integrals.

The goal of our present study is to elucidate the relative accuracy of the PG and DL series as they depend on temperature and density. Because we will span broad parameter ranges and we wish to have accurate values to compare the series against, we choose to study the Lennard-Jones fluid. A general equation of state (EoS) covering both gases and liquids is still a cutting-edge topic in physics, and plethora of EoS have been proposed\cite{Nicolas1979,Karl1993,KOLAFA19941,Sun1996,Boltachev2003,Thol2016,Pieprzyk2018}.  Reference~\cite{Thol2016}, in particular, gives accurate excess entropies
\begin{equation}
  \label{eq:S_ex}
  S^{(ex)}\equiv S^{(EoS)}-S^{(IG)}
\end{equation}
across a broad range of temperature and density, and we shall accept its values as ground truth. In the following, we first introduce the two series formally, then we describe our simulation details, discuss two-body and three-body truncations of the series, and interpolation approaches based on each. Several appendices contain calculational details.

\section{Formal approach}

According to statistical mechanics, the configurational entropy in the canonical ensemble is
\begin{equation}
  \label{eq:S_can}
  S=-\frac{1}{N!\Lambda^{3N}}\int_V \,d{\bm r}^N p({\bm r}^N)\log p({\bm r}^N),
\end{equation}
where $p({\bm r}^N)$ is the N-body probability density function
\begin{equation}
\begin{aligned}
  \label{eq:2}
  p({\bm r}^N) &=\frac{e^{-\beta H({\bm r})}}{Z},\\
  Z &=\frac{1}{N!\Lambda^{3N}}\int_V \,d{\bm r}^Ne^{-\beta H({\bm r})}.
\end{aligned}
\end{equation}
By integrating over a portion of the coordinates, $n$-body correlation functions may be defined~\cite{HSGreen1952,McQuarrie},
\begin{equation}
  \label{eq:gN}
  g_N^{(n)}(\br_1,\dots,\br_n) = \frac{\rho^{N-n}}{(N-n)!}
  \int_V\rmd\br^{(N-n)}~ p(\br^N).
\end{equation}

Following the discussion in Ref.~\cite{Baranyai1989,Widom2019}, entropy may be expressed in two alternate series
\begin{equation}
\begin{aligned}
  \label{eq:s1s2}
  S^{PG} &=s^{PG}_1+s^{PG}_2+s^{PG}_3+\cdots,\\
  S^{DL} &=s^{DL}_1+s^{DL}_2+s^{DL}_3+\cdots,
\end{aligned}
\end{equation}
where $s^{DL}_1$ and $s^{PG}_1$ are the leading terms introduced in Eqs.~(\ref{eq:S1_DL}) and~(\ref{eq:S1_PG}), and the terms that follow
are corrections due to pairs and triplets. In view of the identity~(\ref{eq:expand_1}), the relation between terms of the DL and PG series is
\begin{equation}
  \label{eq:s_n}
  s^{DL}_n=s^{PG}_n+\frac{1}{n(n-1)},\hbox{ for }n>1.
\end{equation}
The two-body term in the dense liquid series is
\begin{equation}
\begin{aligned}
  \label{eq:s2DL}
  s^{DL}_2&=s^{DL}_{2,c} + s^{DL}_{2,i},\\
  s^{DL}_{2,c}&=\frac{1}{2}+\frac{1}{2}\rho\int\rmd\br~(\g{2}(r)-1),\\
  s^{DL}_{2,i}(R)&=-\frac{1}{2}\rho\int\rmd\br~\g{2}(r)\ln{(\g{2}(r))},
\end{aligned}
\end{equation}
and the three-body terms is
\begin{equation}
\begin{aligned}
  \label{eq:s3DL}
  s^{DL}_3 &= s^{DL}_{3,c} + s^{DL}_{3,i},\\
  s^{DL}_{3,c} &= \frac{1}{6}+\frac{1}{6}\rho^2\int\rmd\br^2~
  (\g{3}-3\g{2}\g{2}+3\g{2}-1) \\
  s^{DL}_{3,i} &=-\frac{1}{6}\rho^2\int\rmd\br^2~
  \g{3}\ln{(\g{3}/\g{2}\g{2}\g{2})}.
\end{aligned}
\end{equation}

The DL compressibility terms, $s^{DL}_n$, vanish identically for correlation functions $g_N^{(n)}$ calculated in the canonical ensemble~\cite{HSGreen1952,McQuarrie}, but in the grand canonical ensemble they relate to the isothermal compressibility as~\cite{Mountain1971}
\begin{equation}
\begin{aligned}
  \label{eq:compress}
  s^{DL}_{2,c} &= \frac{1}{2}\gamma,~~~\gamma=\rho\kB T\chi_T,\\
  s^{DL}_{3,c} &= \frac{1}{2}\gamma-\frac{1}{3}\gamma^2 +
  \frac{1}{6}\rho\gamma\frac{\partial\gamma}{\partial\rho}\Bigr|_\beta,
\end{aligned}
\end{equation}
and they represent contributions to the entropy due to density fluctuations. For a perfect gas, $\gamma=1$ and the PG compressibility terms $s^{(PG)}_{n,c}$ vanish, while in a dense liquid, $s^{DL}_{n,c}$ is small but nonvanishing and physically meaningful.

In principle $S^{DL}=S^{PG}$, but their partial sums up to the maximum order $M$,
\begin{equation}
  S^{DL}_M=\sum_{n=1}^M s^{DL}_n,~~~S^{PG}_M=\sum_{n=1}^M s^{PG}_n,
\end{equation}
differ according to
\begin{align}
  \label{eq:S_M}
  S^{PG}_M = S^{DL}_M+\frac{1}{M}.
\end{align}
When the series are truncated at three-body entropy terms $M=3$, the difference between $S_3^{DL}$ and $S_3^{PG}$ gives a constant $\frac{1}{3}$ which is exactly the deviation of calculated entropy from target entropy in Ref.~\cite{Singh2014}, and is likely also the explanation for discrepancies reported in~\cite{CAO20113114,Dhabal2015,Dyre2018}. The advantage of the DL series in the case of nearly incompressible fluids was noted by Laird and Haymet~\cite{Laird1992}. The relation~(\ref{eq:S_M}) makes it convenient to transform calculated results between th DL series and the PG series; consequently our calculations and discussions will mainly focus on the DL series.

\section{Simulation details}
To examine models of both series, we compute their entropies for a Lennard-Jones system and compare our calculations with high quality equation of state (EoS) entropies~\cite{Thol2016}. All Lennard-Jones simulations are performed using LAMMPS~\cite{LAMMPS} in the canonical (NVT) ensemble with mass $m=1.0$, pair potential coefficients $\epsilon=1.0,\sigma=1.0$, cutoff radius $r=2.5$, and a long-range tail correction to the energy. 

In preparation for our entropy calculations, simulations were initialized with $7\times7\times7$ FCC cubic cells (1372  atoms) covering temperatures $T\in[0.7,9]$ and densities $\rho\in[0.1,1.0]$, both in increments of $0.1$. Our runs used a time step of 0.005 and reached a total of $10^6$ steps in each of 36 independent trajectories. After reaching equilibrium, we evaluated two-body and three-body entropy terms for all states in the liquid region of the $\rho-T$ phase diagram. In addition, we carry out simulations of $14\times14\times14$ FCC cubic cells (10976 atoms) at $\rho=1.0$ near the solid-liquid phase transition at $T=1.55$, again using a time step of 0.005, for a total of $1.6\times 10^6$ steps.

See Appendix~\ref{app:A} for details of our evaluations of the required $2$- and $3$-body integrals.

\section{two-body entropy}
\label{sec:2-body}

\begin{figure}[hptb]
  \centering
  \includegraphics[width=.45\textwidth]{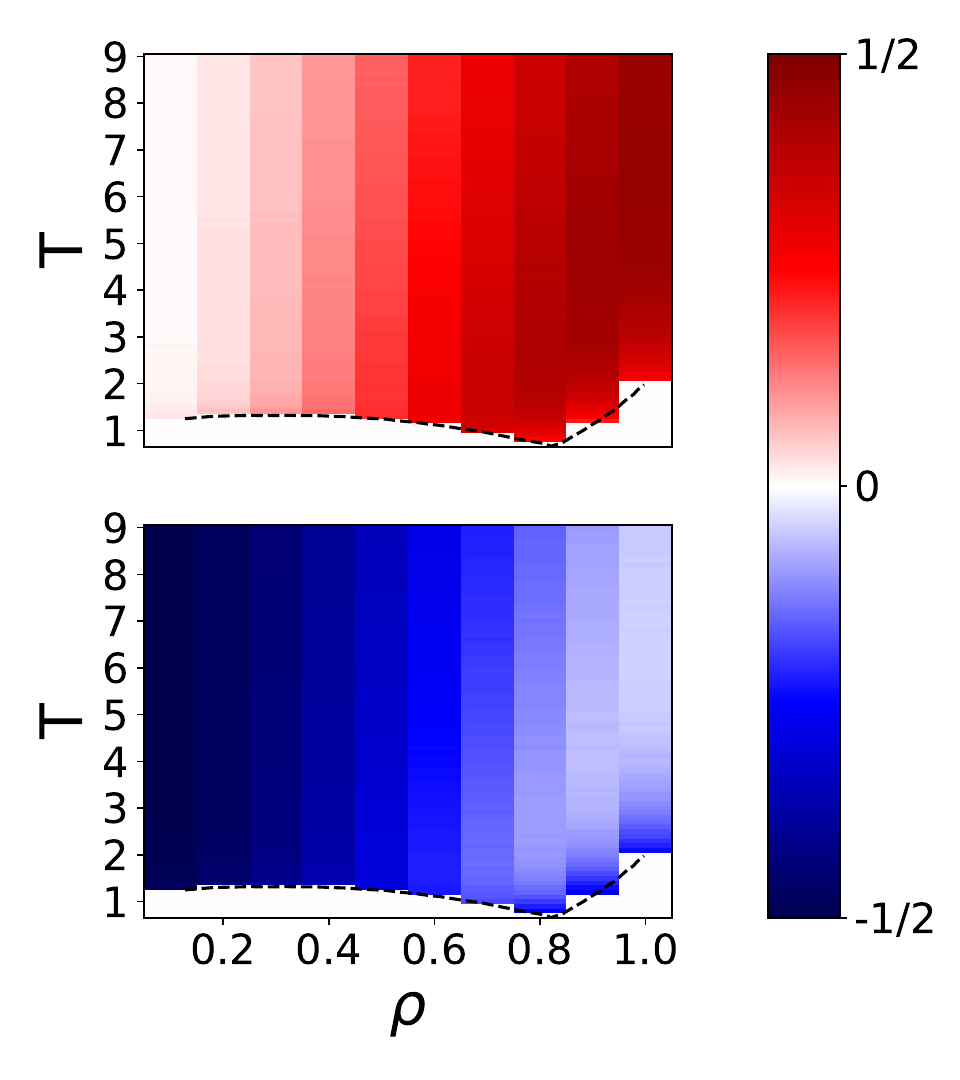}
  \caption{ Residual entropies of PG series, $S_2^{PG}-S^{(EoS)}$ (top) and DL series, $S_2^{DL}-S^{(EoS)}$ (bottom). Black dashed lines denote the liquid-solid phase boundary from Ref.~\cite{Thol2016}.}
  \label{fig:s2si}
\end{figure}

To evaluate the performance of our two series, we truncate the partial sums at the two-body level, $M=2$, for which $S_2^{PG}=S_2^{DL}+1/2$. As shown in Fig.~\ref{fig:s2si}, $S^{PG}$ accurately approaches the known $S^{(EoS)}$ in the limit of low density, as expected. However, the error in $S_2^{PG}$ grows towards $+1/2$ as density rises, and $S_2^{DL}$, which was low by $-1/2$ at low density, becomes increasingly accurate.

\begin{figure}[hptb]
  \centering
  \includegraphics[width=.45\textwidth]{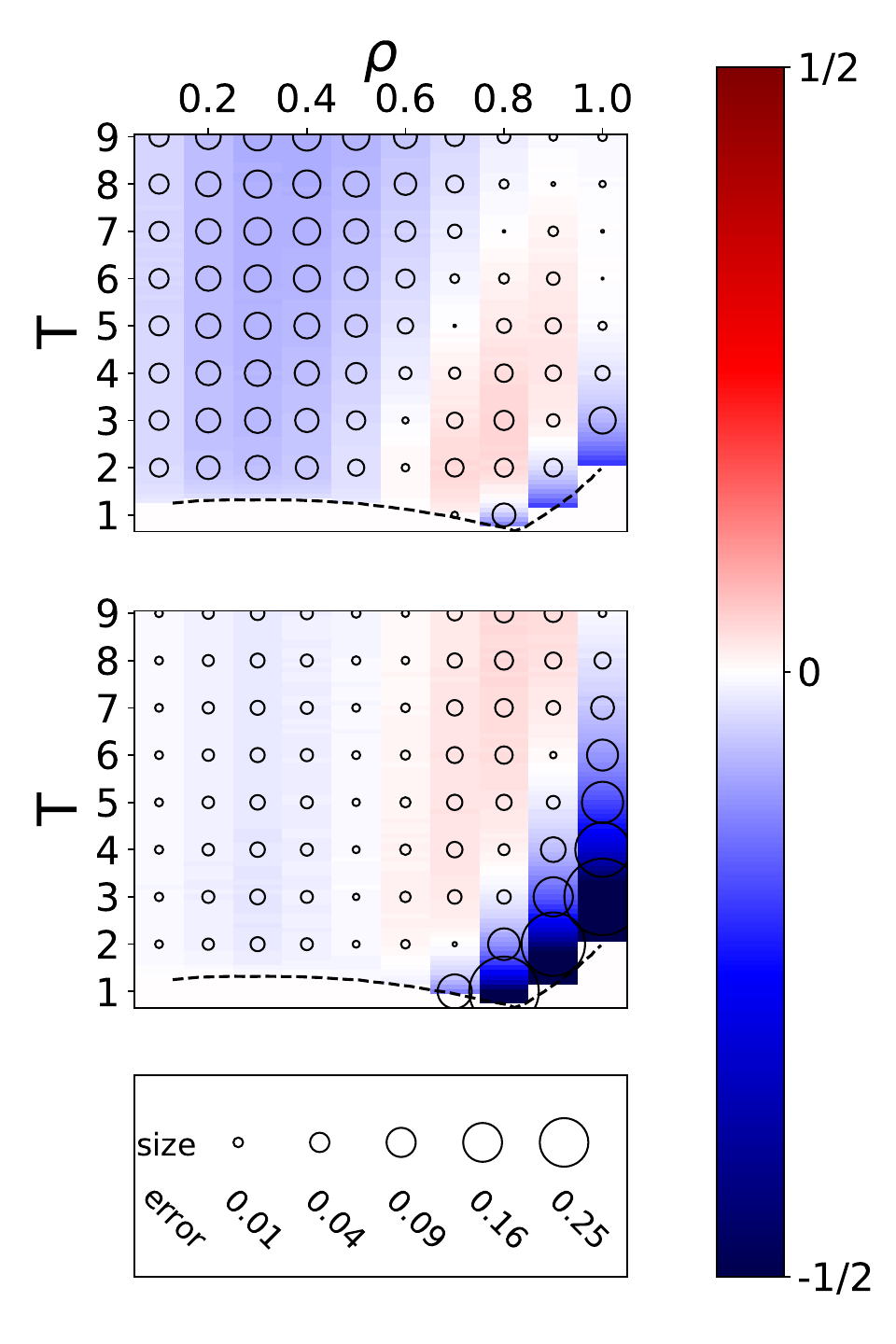}
  \caption{ Residual entropies of empirical interpolations of two series respect to density $\rho$: (top) interpolated entropies $S_2^{(GL)}-S^{(EoS)}$ according to Eq.~(\ref{eq:S_GL}); (bottom) interpolation of $S_\infty-S^{(EoS)}$ according to Eq.~(\ref{eq:S_inf}). Circles denote absolute errors.}
  \label{fig:s2interp}
\end{figure}

We are tempted to try a heuristic interpolation between the two limits; hence we define a formula for a general liquid (GL) as
\begin{equation}
  \label{eq:S_GL}
  S_2^{GL}=\alpha^{DL}S_2^{DL}+\alpha^{PG}S_2^{PG}
\end{equation}
where and $\alpha^{DL}$ and $\alpha^{PG}$ vary with density $\rho$ and satisfy $\alpha^{DL}+\alpha^{PG}=1$. Fig.~\ref{fig:s2interp} (top) shows the result of a linear interpolation with
\begin{equation}
  \label{eq:alpha}
  \alpha^{DL}=\frac{\rho-\rho^{PG}}{\rho^{DL}-\rho^{PG}},\quad
  \alpha^{PG}=\frac{\rho^{DL}-\rho}{\rho^{DL}-\rho^{PG}}
\end{equation}
where we set $\rho^{PG}=0$ and we arbitrarily take $\rho^{DL}=1.1$ as a representative value close to the values near the liquid-solid transition. Our GL entropies are better than either of our initial two series across most of the $\rho-T$ plane.

\section{Three-body entropy}
\begin{figure}[hptb]
  \centering
  \includegraphics[width=.35\textwidth]{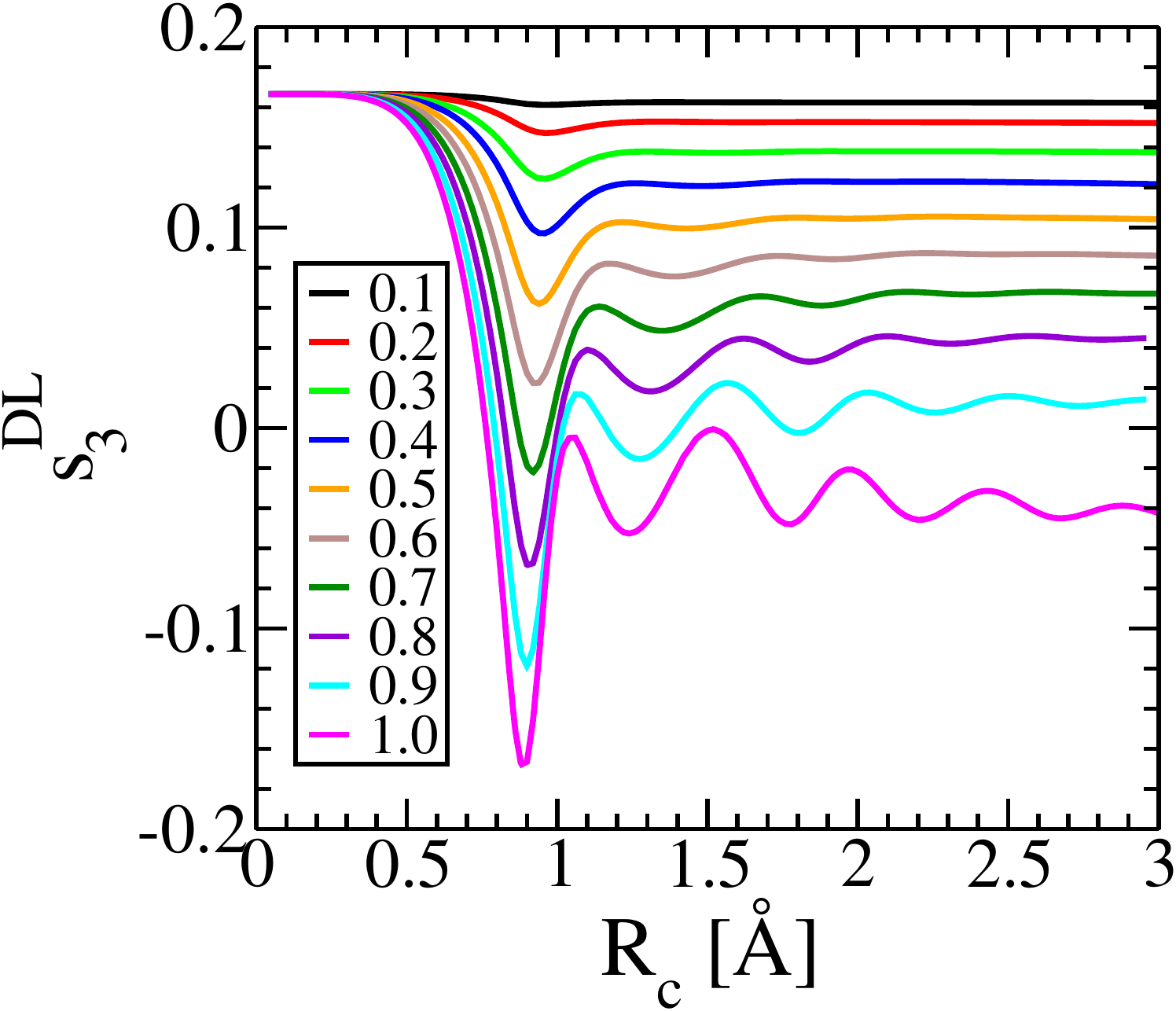}
  \includegraphics[width=.35\textwidth]{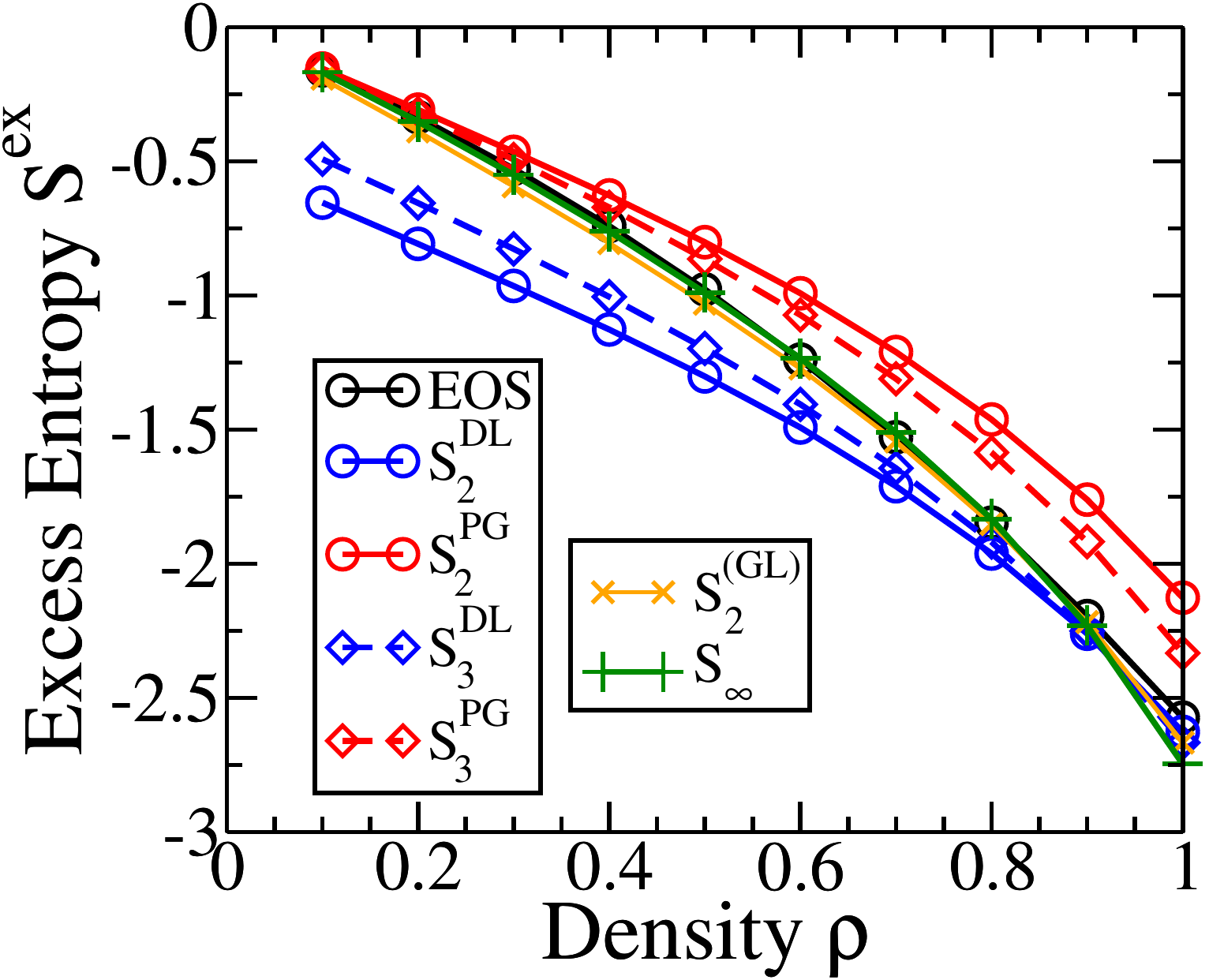}
  \caption{\label{fig:s3}
    Top: Three-body entropies for all densities at $T=5$.
    Bottom: Excess entropies from liquid calculated with EoS,
    $S^{DL}_2$,$S^{PG}_2$, $S^{DL}_3$,$S^{PG}_3$ and interpolated
    entropies according to density $S_2^{(GL)}$ (based on Eq.~(\ref{eq:alpha})) and three-body terms
    $S_\infty$ (based on Eq.~(\ref{eq:S_inf})). }
\end{figure}

In this section, we truncate the series at the three-body terms, $M=3$. We also propose a new interpolation formula based on the three-body entropy that eliminates any ad-hoc parameters (see Fig.~\ref{fig:s2interp}. In Fig.~\ref{fig:s3} we show the three-body term $s_3^{DL}$ at $T=5$ and various densities as a function of integration cut-off radius $R_c$. At low densities $s^{DL}_3$ converges quickly with $R_c$ and tends towards $+1/6$ as $\rho\to 0$. At higher densities the convergence is slower and the values go negative. We also plot the {\em excess} entropy relative to the ideal gas for each of our series as truncated at $M=2$ and $M=3$.

Each $M=3$ partial sum still yields excellent agreement with the EoS entropy in its appropriate density limit. For all densities, $s^{PG}_3$ lowers the excess entropy while $s^{DL}_3$ increases it. As a result, both $M=3$ partial sums ($S^{DL}_3$ and $S^{PG}_3$) are closer to the true EoS than the $M=2$ sums; including the $3$-body terms improves the accuracy of the dense liquid series in the low density limit, and {\em vice-versa}, but evidently additional terms are required for either series to match the known EoS at all densities.

We propose a new method to interpolate between limits that effectively includes higher order terms. Our central assumption is that the ratios of high-order terms in $s^{DL}$ and $s^{PG}$ series are fixed ({\em i.e.} the two series' partial sums converge at the same rate)
\begin{equation}
  \label{eq:ratio}
  \frac{s_3^{PG}}{s_3^{DL}}=\frac{s_4^{PG}}{s_4^{DL}}=...=\frac{s_n^{PG}}{s_n^{DL}}.
\end{equation}
Hence the two fully-summed series satisfy
\begin{equation}
  \label{eq:t}
  t\equiv\frac{s_3^{PG}}{s_3^{DL}}=\frac{S_2^{PG}-S_\infty^{PG}}{S_2^{DL}-S_\infty^{DL}}.
\end{equation}
Because the two series should converge to a common limit,
\begin{equation}
S_\infty^{PG}=S_\infty^{DL}\equiv S_\infty,
\end{equation}
we may solve Eq.~(\ref{eq:t}) for
\begin{equation}
  S_\infty =\frac{tS_2^{DL}-S_2^{PG}}{t-1}.
\end{equation}
Replacing $S_2^{PG}$ with $S_2^{PG}=S_2^{DL}+1/2$ based on Eq.~(\ref{eq:S_M}), $S_\infty$ becomes
\begin{equation}
  S_\infty   =S_2^{DL}-\frac{1}{2(t-1)}.
\end{equation}
Since $t=s_3^{PG}/s_3^{DL}$, and also $s_3^{DL}=s_3^{PG}+1/6$ according to Eq.~(\ref{eq:s_n}), we have
\begin{equation}
\begin{aligned}
  \label{eq:S_inf}
  S_\infty &=S_2^{DL}-\frac{s^{DL}_3}{2(s^{PG}_3-s^{DL}_3)},\\
  &=S_2^{DL}+3s_3^{DL}.
\end{aligned}
\end{equation}
That is, we approximate
\begin{equation}
  \label{eq:S_sum}
  \sum_{n=3}^\infty s_n^{DL}\approx 3 s_3^{DL}.
\end{equation}
This identity holds in the limit $\rho\to 0$, and seems to remain nearly correct for low and moderate density. Applying this formula to the LJ liquid at T=5 as shown in Fig.~\ref{fig:s3}, we find this interpolation model yields excellent agreement with the EoS entropy before ultimately failing at very high density and low temperature.

\begin{figure}[hptb]
  \centering
  \includegraphics[width=.35\textwidth]{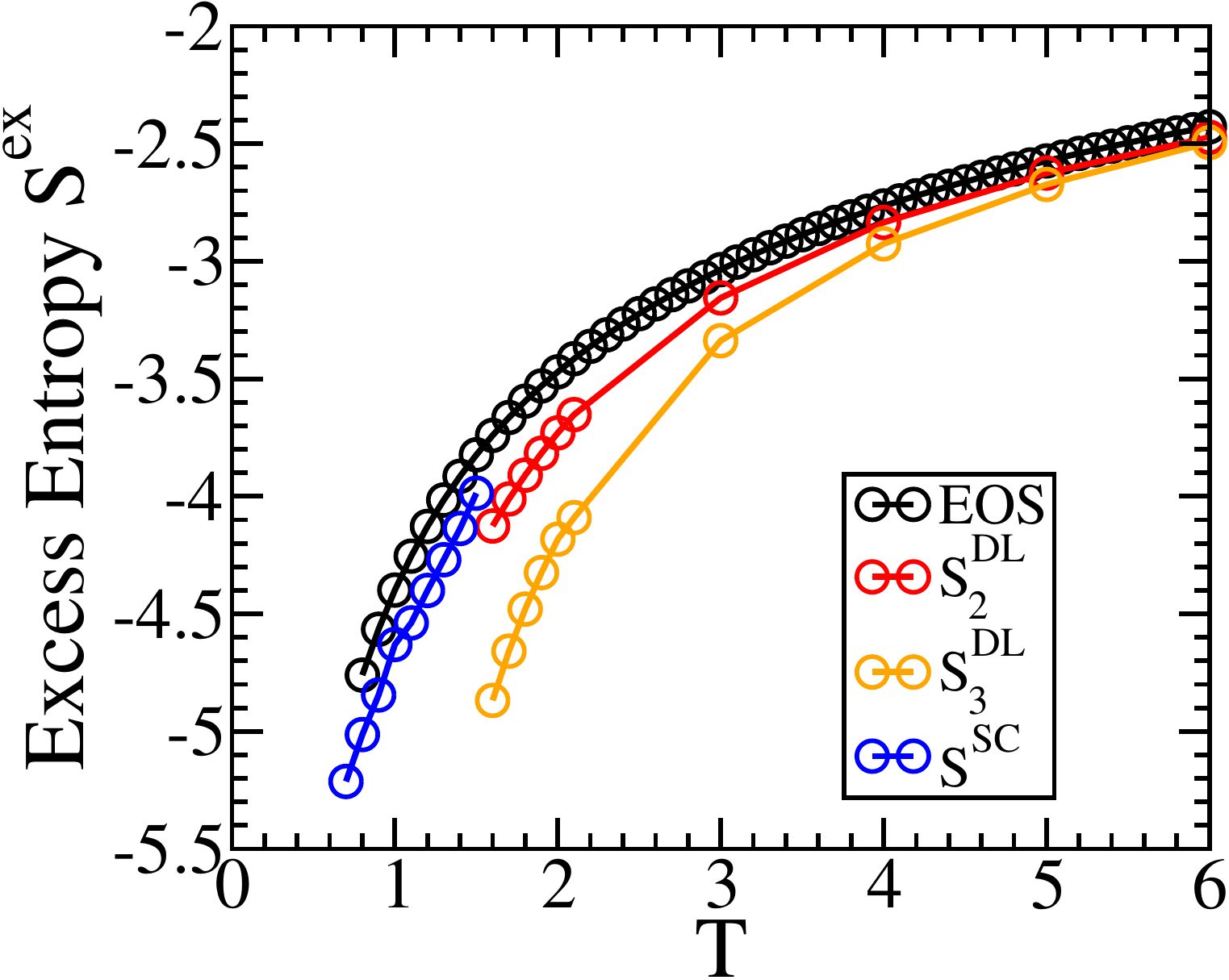}
  \caption{\label{fig:ss}
    Comparison of excess entropies from EoS $S^{\rm EoS}$,  $S^{DL}_2$,$S^{DL}_3$ and solid crystalline entropy $S^{\rm sc}$. }
\end{figure}

We now investigate the failure of our approximations at density $\rho=1.0$, close to the liquid-solid transition. Our calculations across a range of temperature are summarized in Fig.~\ref{fig:ss}. A solid to liquid phase transition occurs at $T_c=1.55$; the solid entropy for $T<T_c$ is calculated using the displacement covariance matrix $\Sigma_{\mathu}$,
\begin{equation}
  \label{eq:S:cov}
  S=\frac{1}{2}\ln\left((2\pi e/\Lambda^2)^{3N}\det{(\bm{\Sigma_\mathu})}\right),
\end{equation}
as described in Ref.~\cite{e24050618}. Here the two-body entropy $S^{DL}_2$ deviates below $S^{EoS}$ as temperature $T$ drops. Including the three-body term $s^{DL}_3$ actually worsens the agreement. We are uncertain if this is genuine, or if instead it is an artifact due to poor convergence of $s_3$ with respect to run time or sample cell size.

\section{Conclusion}

Entropy approximations for liquids geared towards low and high density limits have been tested for the case of Lennard-Jones liquids by comparing to accurate EoS entropies over broad density and temperature ranges. The PG series works best in low density limit, and overestimate entropy in general. The DL series works best in the high density limit, and underestimates entropy in general. Unfortunately, it fails near the solid-liquid phase transition at low temperatures.

The success of our approximations is based on validation of a superposition approximation~\cite{Kirkwood1942} which exclude long range and high order multi-point correlations from the information terms. In this way, all leading terms up to order $M$ are included,
\begin{equation}
  S=s_1+\sum_{k=2}^M s_k+\sum_{k=M+1}^\infty s_{k,c},
\end{equation}
but higher order multi-point information terms $s_{k,i}$ vanish, leaving only compressibility contributions~\cite{Raveche1971,Mountain1971}. Expressions for $s_{2,c}^{DL}$ and $s_{3,c}^{DL}$ are proportional to the isothermal compressibility as shown in Eq.~(\ref{eq:compress}). Compressibility drops as density grows, explaining why the DL series is preferred at high density. In contrast, the compressibility terms of the PG series vanish at low density and approach $-1/n(n-1)$ at high density, explaining why the PG series is accurate at low density but overestimates entropy at high density.

We quantify the relative accuracy of the PG and DL series at the two- and three-body levels, revealing the need for interpolation methods that bridge the two series between their respective density limits. Two different interpolation schemes are proposed: one of them (Eq.~(\ref{eq:S_GL})) depends upon an ad-hoc parameter that we call $\rho^{DL}$ at which the dense liquid series is (incorrectly) presumed exact; the other (Eq.~(\ref{eq:S_inf})) is parameter-free but presumes validity of an approximate relationship (Eq.~(\ref{eq:ratio})) between $n$-body terms in each series. Both series ultimately fail at the very highest densities.

The excess entropy has been related to dynamic properties of liquids such as their diffusion coefficients by the excess entropy scaling rule~\cite{Rosenfeld1977,Dzugutov1996,Dyre2018}. It is recommended to adopt a more accurate excess instead of the two-body form applied in Dzugutov's law~\cite{Dzugutov1996}, in order to justify a universal excess entropy scaling law in liquid metals~\cite{Hoyt2000,Jakse2016}. Our interpolation methods may provide better entropy approximations for many other purposes, such as prediction of melting points, liquid-liquid phase separation, and eutectics.

\section{Acknowledgments}
This work was supported by the Department of Energy under grant DE-SC0014506. This research also use d the resources of the National Energy Research Scientific Computing Center (NERSC), a US Department of Energy Office of Science User Facility operated under contract number DE-AC02-05CH11231 using NERSC award BES-ERCAP24744.

\appendix
\section{Numerical evaluation of $n$-body integrals}
\label{app:A}

The appendix discusses details of data analysis and the evaluation of integrals in Eqs.~(\ref{eq:s2DL}) and~(\ref{eq:s3DL}). Relevant codes may be found at~\cite{utilities}.

\subsection{Representation and integration of $\g{2}$}
\label{app:rep-g2}

Two-body correlation functions are accumulated using histograms with a bin width of 0.01 without any smearing. Smearing would reduce the amplitudes of oscillations and tends to reduce the magnitude of the information terms owing to the concavity of the logarithm. However, random sampling noise introduces false information that tends to enhance the magnitude of information terms. We counter this effect through the use of long run times supplemented with an extrapolation to the limit of infinite time.

The histogram is normalized by
  \begin{equation}
    f(R) = \frac{\Delta N(R)}{V\Delta V \rho_2}
  \end{equation}
  where $\Delta N(R)$ is the number of pairs within the spherical shell $r\in [R,R+\Delta R]$ and $\Delta V=4\pi[(R+\Delta R)^3 - R^3]$ is the volume of the shell. The pair density
  \begin{equation}
    \rho_2 = \frac{N(N-1)}{2V^2}
  \end{equation}
with a total number of atoms $N$, box volume $V$, and a factor of $1/2$ to avoid double counting. We then multiply by $4\pi r^2$ and integrate from the first nonzero value up to a variable cutoff distance $R_c$ through trapezoid integration.


\subsection{Representation of $\g{3}$}
\label{app:rep-g3}
\begin{figure}[hptb]
  \centering
  \includegraphics[width=.35\textwidth]{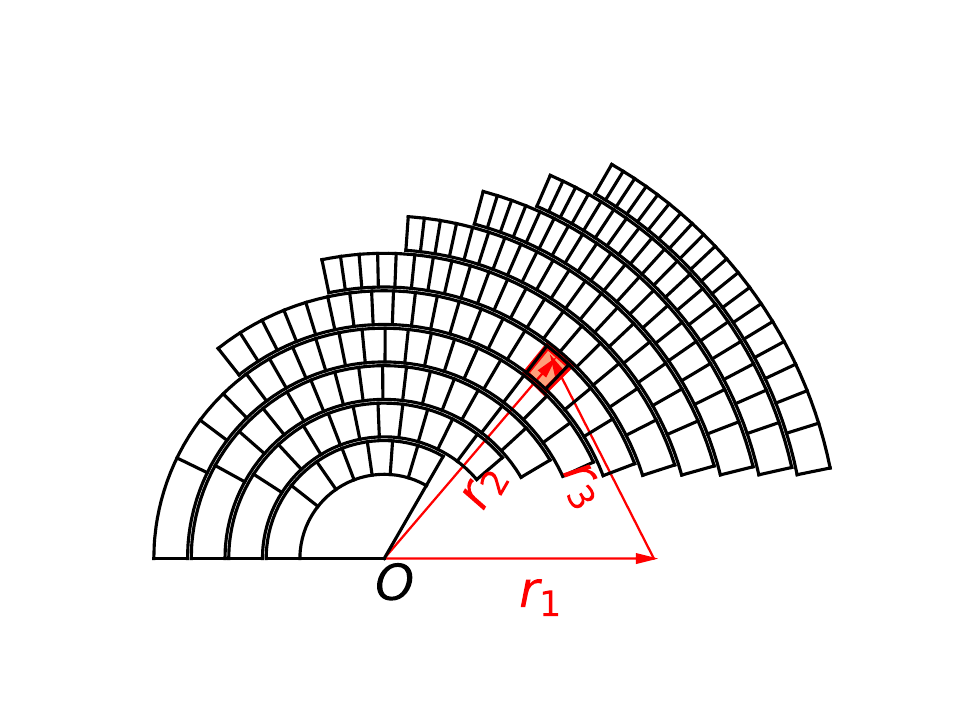}
  \caption{\label{fig:binshape}
    A sketch showing different bin shapes at different values of  $r_1,r_3$ (given $r_2=r_1$) in our binning method.}
\end{figure}

In a uniform and isotropic liquid, the triplet correlation function depends only on the magnitudes of the particle separations $r_1, r_2, r_3$, or equivalently on two separations $r1, r_2$, and $t=\cos{\theta}$, with $\theta$ the angle between bonds $r_1$ and $r_2$. We choose to represent $\g{3}(r_1,r_2,\theta$) as a histogram using bins as illustrated in Fig.~\ref{fig:binshape}.

For a given minimum inner radius $R_{min}$, outer cutoff radius $R_c$, and radial interval $\Delta R$,
\begin{equation}
  r_1,r_2\in \{R_{min}+k \Delta R, k=0,1,..., N_R\}
\end{equation}
where $N_R$ is the number of bins in $R$
\begin{equation}
  N_R=[\frac{R_c-R_{min}}{\Delta R}].
\end{equation}
Angular binning satisfies
\begin{equation}
  \begin{aligned}
  t_{min}&=\max\left[\frac{r_1^2+r_2^2-R_c^2}{2r_1r_2},-1\right],\\
  t_{max}&=\min\left[\frac{r_1^2+r_2^2-R_{min}^2}{2r_1r_2},1\right]
\end{aligned}
\end{equation}
and
\begin{equation}
\begin{aligned}
  N_t&=2\left[\frac{R_{3,max}-R_{3,min}}{\Delta R}\right];\\
  R_{3,min}&=max\left[R_{min},r_2-r_1-\Delta R\right];\\
  R_{3,max}&=min\left[R_{c},r_1+r_2+2\Delta R\right];
\end{aligned}
\end{equation}
supposing $r_2>r_1$.

In this framework, the volume element is
\begin{align}
  \Delta V&=8\pi^2\int_{R_1}^{R_1+\Delta R}\int_{R_2}^{R_2+\Delta
  R}\int_{t}^{t+\Delta t}r_1^2r_2^2\,dr_1\,dr_2\rmd\tau\nonumber\\
  &=\frac{8\pi^2}{9}\left[(R_1+\Delta
    R)^3-R_1^3\right]\left[(R_2+\Delta R)^3-R_2^3\right]\Delta t.
\end{align}
The histogram is normalized with
\begin{equation}
  f(R_1,R_2,t)=\frac{N(R_1,R_2,t)}{V\Delta V(R_1,R_2,t)\rho_3}
\end{equation}
where $N(R_1,R_2,t)$ is the number of triplets that satisfy
$R_1<r_1<R_1+\Delta R$, $R_2<r_2<R_2+\Delta R$, and $t<\tau<t+\Delta
t$. Triplet density $\rho_3$ is defined by
\begin{equation}
  \rho_3=\frac{N(N-1)(N-2)}{V^3}.
\end{equation}

\subsection{Integration of $\g{3}$}
\label{app:int-g3}
The three-body term has two separate contributions, $s^{DL}_{3,c}$ and $s^{DL}_{3,i}$,
\begin{align}
  \label{eq:3b}
  s^{DL}_3&=s^{DL}_{3,c}+s^{DL}_{3,i};\nonumber\\
  s^{DL}_{3,c}&=\frac{1}{6}+\frac{1}{6}\rho^2\int\rmd\br^2[g^{(3)}-3g^{(2)}g^{(2)}+3g^{(2)}-1];\nonumber\\
  s^{DL}_{3,i}&=-\frac{1}{6}\rho^2\int\rmd\br g^{(3)}\ln g^{(3)}/g^{(2)}g^{(2)}g^{(2)}.
\end{align}
Following Ref.~\cite{Mountain1971}, we split the compressibility term, $s^{DL}_{3,c}$ into two terms, $A$ and $B$, and also the constant $\frac{1}{6}$, by subtracting
$g^{(2)}g^{(2)}g^{(2)}$ from $\g{3}$ in $A$ and adding it back into $B$
\begin{equation}
  \label{eq:split_AB}
\begin{aligned}
  s^{DL}_{3,c}&=\frac{1}{6}+A+B; \\
  A&=\frac{1}{6}\rho^2\int\rmd\br^2(g^{(3)}-g^{(2)}g^{(2)}g^{(2)}); \\
  B&=\frac{1}{6}\rho^2\int\rmd\br^2(g^{(2)}-1)(g^{(2)}-1)(g^{(2)}-1),
\end{aligned}
\end{equation}
which can be verified with the identity
\begin{equation}
  \label{eq:supp}
  \Pi(g^{(2)}-1)=g^{(2)}g^{(2)}g^{(2)}-3g^{(2)}g^{(2)}+3g^{(2)}-1.
\end{equation}
Kirkwood's superposition approximation~\cite{Kirkwood1942} claims
\begin{equation}
  \label{eq:sp}
  \lim_{r_1,r_2,r_3->\infty}g^{(2)}(r_1)g^{(2)}(r_2)g^{(2)}(r_3)=g^{(3)}(r_1,r_2,r_3),
\end{equation}
in which case the integrands in $A$ and in $s^{DL}_{3,i}$ both vanish.

Another advantage of this framework is that $\g{2}$ and $\g{3}$ both vanish whenever any of their arguments $r$ falls below a value $R_{min}\approx 0.8$, leaving only the integrand of $B$ nonzero. Since the integrand of B involves two-body correlation functions only, which are more accurate than the 3-body distribution function due to numerical statistics, $B$ can be calculated with high precision. It also saves tremendous work by only requiring $R_{min}<r_1<R_{max}$ and $R_{min}<r_2<R_{max}$ when computing 3-body distribution functions.

\subsection{Convergence analysis}
\label{app:converge-g3}

In principle, the integrals $A$ and $B$, and hence $s^{DL}_{3,i}$, converge when two condition are satisfied: the far field approximation
\begin{equation}
  \label{eq:ffa}
  \lim_{r->\infty} g^{(2)}(r)=1,
\end{equation}
where the distribution of particles becomes uniform; and the superposition approximation Eq.~(\ref{eq:supp}), where the probability of a triplet factors into probabilities of its three pairs. In this section, we briefly discuss the speed of convergence by order analysis.

First examine the far field approximation in the two-body entropy. The integrand of $S_2$ is
\begin{equation}
  \label{eq:s2i}
  f(r)=g(r)-1-g(r)\ln g(r).
\end{equation}
Let $g(r)=1+\delta (r)$, with $\delta (r)$ a decaying oscillation. Substituting this $g(r)$ into Eq.~(\ref{eq:s2i}) and expanding to second order, we have
\begin{equation}
  \label{eq:od}
  f=\delta (r)-[1+\delta (r)][\delta(r)-\frac{1}{2}\delta^2(r)]=-\frac{1}{2}\delta^2(r).
\end{equation}
The oscillatory terms in $S_{2,i}^{DL}$ and $S_{2,c}^{DL}$ cancel leaving a
quadratic term $-\frac{1}{2}\delta^2(r)$. The integral of $f$ up to $R_c$ decreases monotonically with $R_c$, as seen in Fig.~\ref{fig:g2}(b). It convergences provided that $\delta^2(r)$ falls off faster than $1/r^3$.

\vspace{5pt}\vspace{1in}
\label{sec:convg}
\begin{figure}
  \centering
  \includegraphics[width=.45\textwidth]{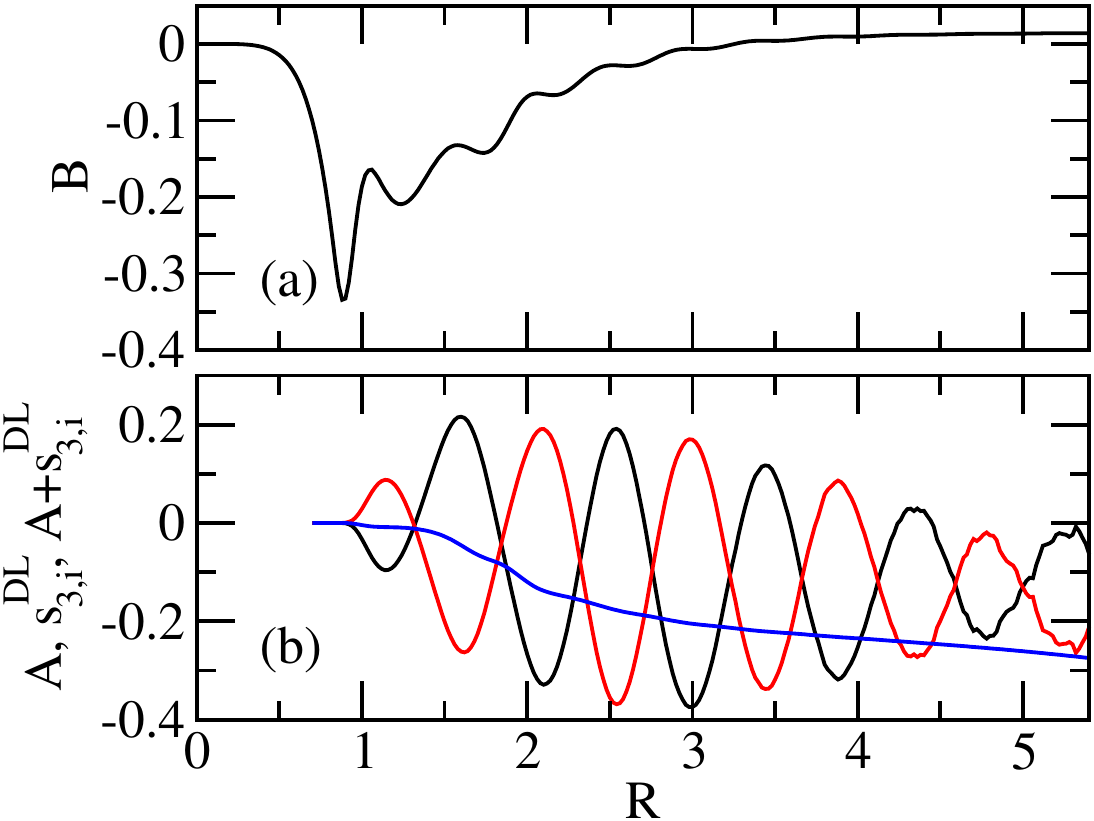}
  \caption{\label{fig:convg} (a) integral $B$ from $s_{3,c}^{DL}$ as defined in Eq.~(\ref{eq:split_AB}). (b) Integrals $A$ (black), $s^{DL}_{3,i}$(red) and $A+s^{DL}_{3,i}$(blue).}
\end{figure}

In the three-body compressibility term, the far field approximation only influences the convergence of integral $B$, with integrand
\begin{equation}
  b = \delta(r_1)\delta(r_2)\delta(r_3).
\end{equation}
Its integration to radius $R_c$ is plotted in Fig.~\ref{fig:convg}(a) which reaches a steady plateau as $g(R)$ approaches to $1$.

The sum of $A$ and $s^{DL}_{3,i}$ has the integrand
\begin{equation}
  h=g^{(3)}-g^{(2)}g^{(2)}g^{(2)}-g^{(3)}\ln g^{(3)}/g^{(2)}g^{(2)}g^{(2)}.
\end{equation}
In the superposition limit
\begin{equation}
  g^{(3)}=g^{(2)}g^{(2)}g^{(2)}+\epsilon(r_1,r_2,r_3),
\end{equation}
and
\begin{equation}
  h\simeq -\frac{1}{2}\epsilon^2(r_1,r_2,r_3).
\end{equation}

Convergence of $B$ requires that $\epsilon(r_1,r_2,r_3)$ falls off faster than $1/r_1r_2r_3$. Also, convergence of $A$ requires that $\delta(r_1)\delta(r_2)\delta(r_3)$ falls off faster than $1/r_1^2r_2^2r_3^2$, {\em i.e.} $|\delta(r)|<1/r^2$. Thus, convergence of $s_3$ imposes stricter bounds than does convergence of $s_2$.

\begin{figure}[hptb]
  \centering
  \includegraphics[width=.35\textwidth]{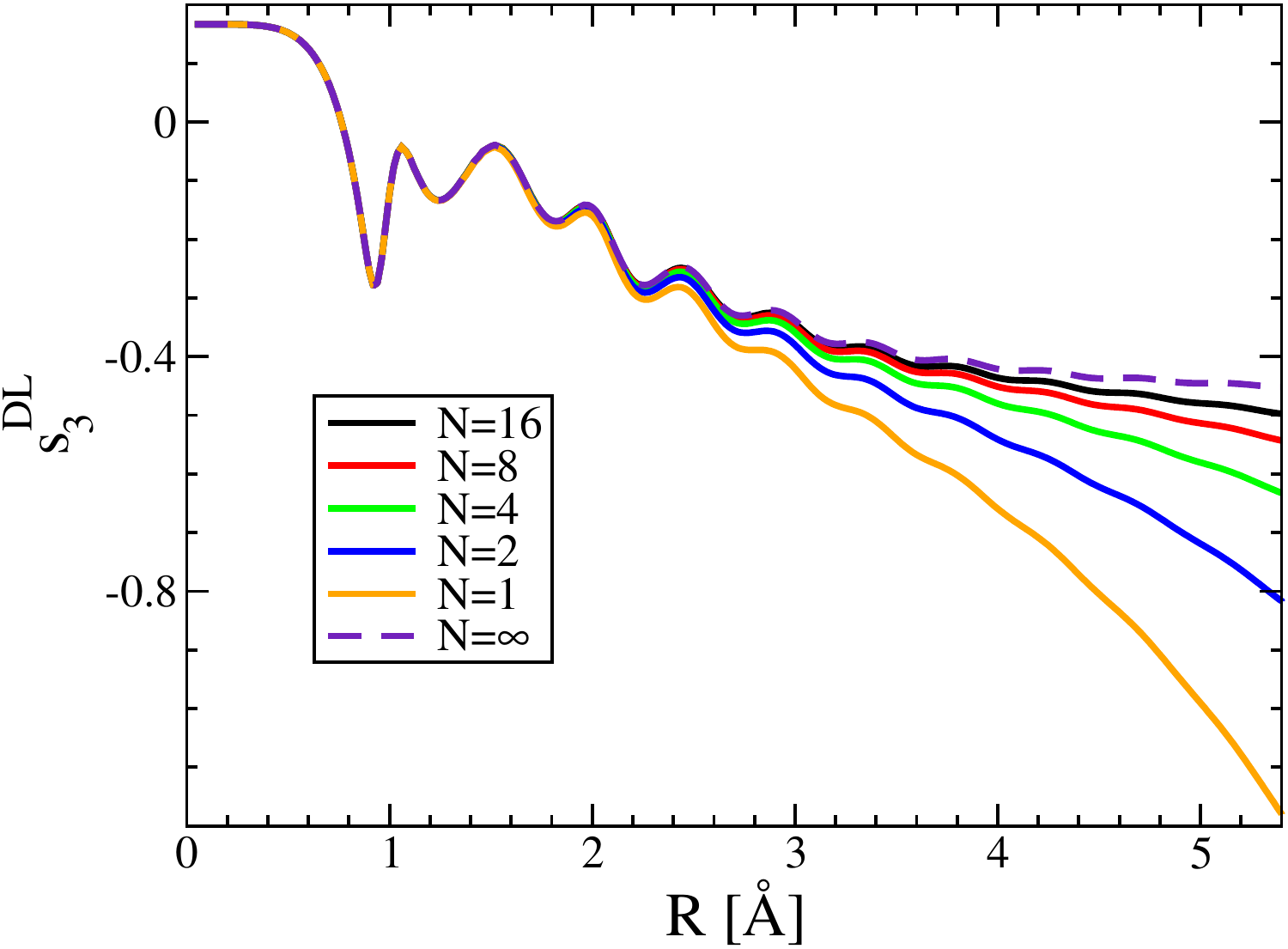}
  \caption{\label{fig:s3n}
    Convergence of three-body entropy $s^{DL}_3$ with respect to number of samples $N\times 10^3$ included in computing three-body distribution function for 10976 atoms at $\rho=1.0$ and $T=2.0$. The Richardson extrapolation is drawn in a dashed line.}
\end{figure}

$S_3$ should decrease monotonically with $R_c$. In Fig.~\ref{fig:convg}(b), the sum of $A$ and $s^{DL}_{3,i}$ seems to converge near $R\sim 3$ but then continues to decrease past $R=5$. This behavior is an artifact due to statistical noise in the three-body distribution function histogram. The noise vanishes at a rate of $1/N$, where $N$ is the number of independent sampled configurations~\cite{Baranyai1990,Luca2019}. Fig.~\ref{fig:s3n} shows examples of divergent $s^{DL}_3$ due to insufficient sampling. With more structures included, the tail of $s^{DL}_3$ turns up and gradually converges. We apply Richardson extrapolation~\cite{richardson1911ix} to estimate the limit of long run time, $s_3(N=\infty)\approx 2s_3(2N)-s_3(N)$.

\bibliography{ref}
\end{document}